\pdfoutput=1
\documentclass{JINST}
\usepackage{graphicx}
\usepackage{rotating}
\usepackage{supertabular}
\usepackage{longtable}
\usepackage{lscape}
\usepackage{multicol}
\usepackage{multirow}
\usepackage{bm}
\long\def\symbolfootnote[#1]#2{\begingroup\def\thefootnote{\fnsymbol{footnote}}\footnote[#1]{#2}\endgroup}

\title{Gator: a low-background counting facility at the \\ Gran Sasso Underground Laboratory}

\author{L. Baudis$^a$\thanks{Corresponding author.}, A.D. Ferella$^a$,  A. Askin$^a$, J. Angle$^b$, E. Aprile$^c$, T. Bruch$^a$, A.Kish$^a$, M. Laubenstein$^d$, A. Manalaysay$^a$, T.~Marrod\'an~Undagoitia$^a$~ and M. Schumann$^a$\\
\llap{$^a$}Physics Institute,\\
University of Z\"urich, Winterthurerstrasse 190, CH-8057, Z\"urich, Switzerland\\
\llap{$^b$}Department of Physics,\\
University of Florida, Gainesville, FL 32611, USA\\
\llap{$^c$}Department of Physics,\\
 Columbia University, New York, NY 10027, USA\\
\llap{$^c$}Gran Sasso National Laboratory,\\
  Assergi, L'Aquila, 67010, Italy\\
  E-mail: \email{laura.baudis@physik.uzh.ch}}

\abstract
{A low-background germanium spectrometer has been installed and is being operated in an ultra-low background 
shield (the Gator facility) at the Gran Sasso underground laboratory in Italy (LNGS). With an integrated rate of $\sim$0.16 events/min in the energy 
range between 100-2700\,keV, the background is comparable to those of the world's most sensitive germanium detectors. 
After a detailed description of the facility, its background sources as well as the calibration and efficiency measurements are introduced.
Two independent analysis methods are described and compared using examples from selected sample measurements. 
The Gator facility is used to screen materials for XENON, GERDA, and in the context of next-generation astroparticle physics facilities such as DARWIN.}


\begin{document}

\section{Introduction}

Gamma-ray spectroscopy offers a standard method for material screening and selection for rare-event 
searches, such as the direct detection of Weakly Interacting Massive Particles (WIMPs)\,\cite{Jungman:1995df} or the 
search for the neutrinoless double beta decay\,\cite{Heusser:1995wd}. An ultra-low level germanium spectrometer in 
a dedicated low-background shield (the Gator facility) has been built and installed at the Laboratori Nazionali del Gran Sasso (LNGS), Italy. 
While the facility is being operated mainly in the context of the XENON program\,\cite{Angle:2007uj}\cite{Aprile:2009yh}\cite{Aprile:2010um} \cite{Aprile:2011vb}, recently it also has been used  for 
the GERDA experiment\,\cite{Gerda_Exp} as well as for R\&D purposes within low-background particle astrophysics.

This paper is organized as follows: in the next section, the Gator facility is described in detail. In 
the third section,  calibration measurements, 
efficiency determinations for the different sample geometries, and cross-checks with standard, calibrated samples are described.
 In the fourth section,  two different data analysis methods are introduced. The fifth section describes the determination of the main background sources of the facility 
 as well as selected results for sample measurements. In the final section, a summary and discussion of future plans  are presented.

\section{The Gator facility}

\begin{figure}[h!]
\includegraphics[scale=0.42]{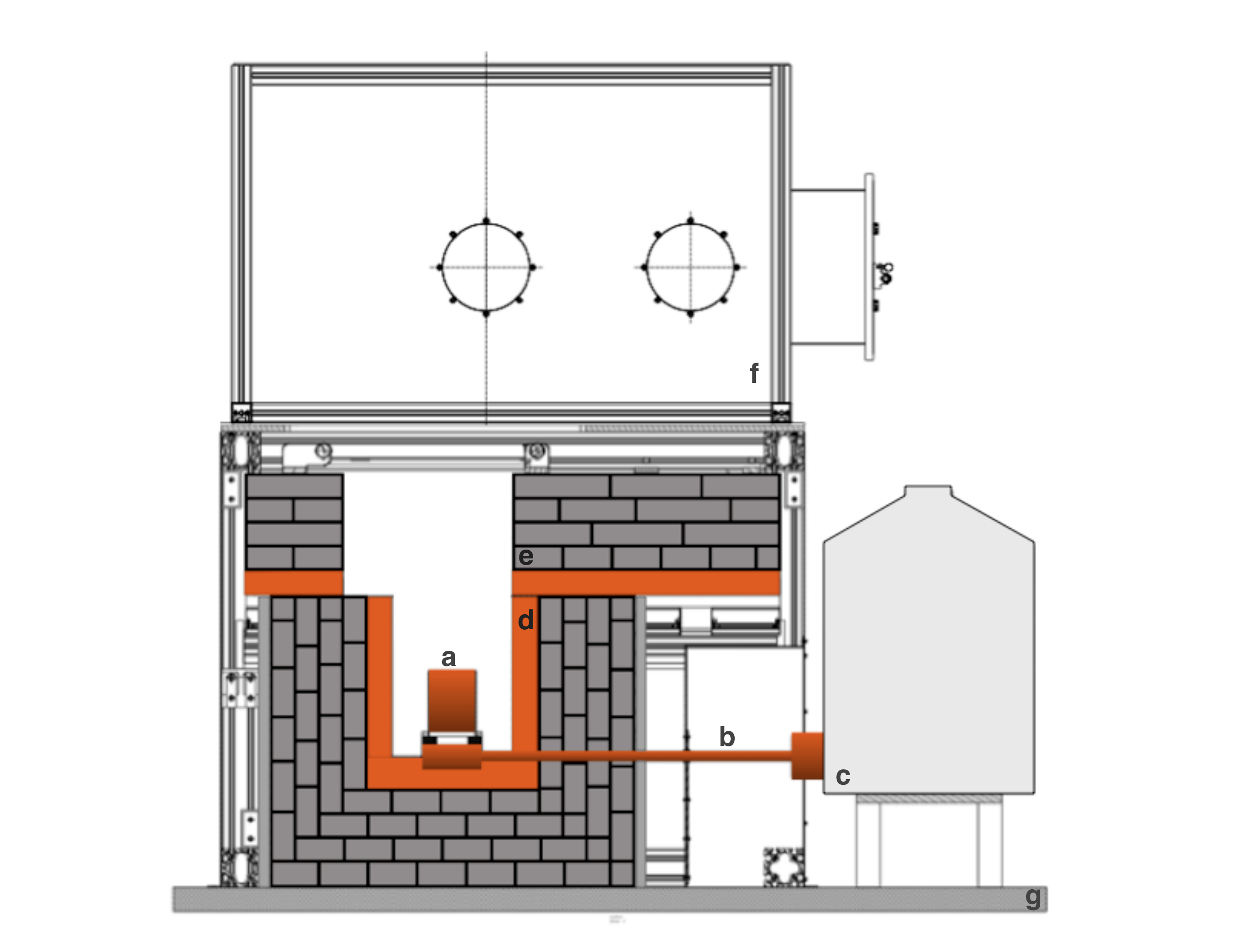}
\centering
\caption{\small{Schematic figure of the Gator facility at LNGS. The HPGe detector (a) with its cold finger (b) and dewar (c), the open sample chamber,  the copper (d) and lead (e) shield with the sliding door, the glove 
box (f) and polyethylene shield (g) can be seen.}}
\label{fig:cutview}
\end{figure}

The design of the facility has been inspired by the layout of the world's most sensitive germanium spectrometers, operated at 
LNGS in connection with the Borexino and GERDA experiments\,\cite{neder2000}\cite{Heusser:2004}.
The core  consists of  a high-purity, p-type coaxial germanium (HPGe) detector with 2.2\,kg of  sensitive
mass with a relative efficiency of 100.5\%\footnote{The quoted efficiency is defined relative to a 7.62\,cm $\times$ 7.62\,cm NaI(Tl) crystal, for the 1.33\,MeV 
$^{60}$Co photo-peak, at a source-detector distance of 25\,cm\,\cite{Knoll}.}. The  detector construction has been performed in close cooperation 
 with Canberra semiconductors\,\cite{Camberra}, using only materials  with ultra-low intrinsic 
radioactive contamination.  The cryostat is made of ultra-low activity, oxygen-free copper  with the cooling
provided by a copper dipstick ('cold finger') in thermal contact with a liquid nitrogen bath. The  cryostat is of U-type, with the cooling 
rod shaped in a right angle below the cryostat to avoid direct line-of-sight to the outside (see Figure \ref{fig:cutview}). While the field-effect transistor (FET) used for the charge readout  
is cooled and close to the detector, the preamplifier is placed outside the low-background shield, since it contains more radioactive components.

The shield of the detector has been designed to provide a large sample capacity, an ultra-low background and easy access 
to the germanium spectrometer itself. The sample chamber, with a dimension of 25$\times$25$\times$33\,cm$^3$, is surrounded by 5\,cm 
(7\,cm for the base plate) of oxygen-free, radio-pure copper from Norddeutsche Affinerie\,\cite{Aurubis}. Residual surface contaminations  
of the copper plates were removed by treating them with diluted sulfuric and citric acid solutions, followed by cleaning 
with deionized water. All steps were performed under clean-room conditions.  
The copper is surrounded by four layers of lead from Plombum\,\cite{Plombum},  each 5\,cm thick. The inner 5\,cm lead layer has a 
 nominal $^{210}$Pb activity of 3\,Bq/kg, while the outer 3 layers were built from lead with a higher  $^{210}$Pb activity of
75\,Bq/kg. All lead bricks were cleaned with ethanol before being installed in the shield. Their 
arrangement is such that no direct line-of-sight to the HPGe detector is possible. Large copper plates, which 
close the sample cavity and carry the upper lead bricks, were placed on sliding rails allowing for easy opening and closing of the shield: 
The full sample chamber can be accessed when the top shield is open. 
The lead is surrounded by 5\,cm of polyethylene as a shield against ambient neutrons, and the entire shield is enclosed 
in an airtight aluminum box in order to prevent radon entering the system.

Figure \ref{fig:cutview} shows a schematic view of the detector and  shield configuration.  A sample handling and 
glove box made of plexiglass, including an airlock system,  is placed on top of the aluminum housing. 
The entire system is continuously flushed at slight over-pressure with boil-off nitrogen gas  to 
suppress radon diffusion into the shield. A  5\,mm inner diameter PTFE tube allows sealed calibration sources 
to be brought close to the germanium detector.  The samples to be screened are first cleaned in an ultra-sonic bath 
of ethanol (where applicable), then enclosed in a sealed plastic bag for transportation to the underground site, after which they are stored for a few days under nitrogen atmosphere above the closed chamber.  
This allows trapped radon and plate-out $^{220}$Rn and $^{222}$Rn progenies to decay before the actual counting starts.

To remotely check the stability of the detector with time, a monitor system has been installed. The liquid nitrogen level, the flow of nitrogen gas, the leakage current 
across the germanium diode and the high-voltage applied to the diode are read every 2 minutes, while the overall trigger rate is read every 6 hours. 
The liquid nitrogen level is measured with a  level-meter consisting of two 40\,cm long concentric 
metal tubes acting as a capacitor whose capacitance is read out with a 
universal transducer interface board. The nitrogen gas flow into the shield is monitored  
and regulated with an electronic flowmeter and the leakage current is measured as the voltage drop across the feed-back resistor and read out with an analog-to-digital converter. 
The relevant parameters are plotted versus time and can be monitored on a web interface which is refreshed every 10 minutes. In case pre-defined thresholds for these 
parameters are not met, email and SMS alarms are being issued. The data acquisition consists of a high-voltage unit, a spectroscopy amplifier, and a a self triggering, 16\,k channel 
multichannel analyzer. The spectra, acquired and cleared every 6 hours, are saved in ascii files and analyzed offline.

Prior to its installation at LNGS, the germanium spectrometer was  operated in the Soudan underground laboratory in northern Minnesota, within the SOLO facility\,\cite{SOLO}. 
At the Soudan laboratory, the detector was used for the screening of XENON10\,\cite{Angle:2007uj} materials, and several background runs were acquired\,\cite{AnglePhD}. 
During its water- and ground-based transportation to LNGS, the detector was  exposed for several months to the cosmic ray flux at the Earth's surface, leading to cosmic 
activation of the crystal and the surrounding copper of the cryostat.  The shield was improved in October 2008, and it was cleaned once again in February 2009.

Table \ref{tab:BGintegral} shows the integral background in the 100-2700\,keV region in the Soudan configuration, and for three measurements taken at LNGS. 
The integral counting rate at LNGS has been constantly  decreasing  due the decay of cosmogenic radio-nuclides  such as $^{54}$Mn,  $^{57}$Co, $^{58}$Co and $^{65}$Zn, with typical half-lives around one year on the one hand, and due to  an improved shield and overall sealing of the system on the other hand. This is reflected in the decrease of prominent lines from $^{214}$Bi and $^{214}$Pb, as shown in Table 
\ref{tab:BGindividual}.  It gives the integral counting rates in the $\pm$3$\sigma$-regions for the main primordial $\gamma$-lines, as well as for the  $^{137}$Cs, $^{60}$Co and 
$^{40}$K lines, along with a comparison with the GeMPI detector, which is one of the world's most sensitive low-background spectrometers\,\cite{neder2000}. 

 \begin{table}[h]
\begin{center}
\caption{\label{tab:BGintegral}\small{Integral background counting rates for Gator as
measured at Soudan and at LNGS in three different runs. The integral is evaluated in the energy range [100, 2700] keV. }}
\vspace{0.3cm}
\begin{tabular}{|l|p{3.5cm}p{4cm}|}
\hline
Run   & Lifetime [days] & Rate [events/min]\\
\hline
Gator at Soudan		&   22.96	& 0.842 $\pm$ 0.005 \\
Gator at LNGS (09-2007)	&   14.90	& 0.258 $\pm$ 0.003 \\
Gator at LNGS (10-2008)	&   22.59	& 0.186 $\pm$ 0.003 \\
Gator at LNGS (04-2010)	&   51.43	& 0.157 $\pm$ 0.001 \\
\hline
\end{tabular}
\end{center}
\end{table}

\begin{table}[h]
\begin{center}
\caption{\label{tab:BGindividual}
\small{Background counting rates (in events/day) in the $\pm$3$\sigma$-regions for the main primordial and the gamma lines of $^{137}$Cs, $^{60}$Co and  $^{40}$K.}}
\vspace{0.3cm}
\begin{tabular}{|r|l|p{2cm}p{2cm}p{2cm}p{2cm}c|}
\hline
Energy & Chain/nuclide	& \multicolumn{5}{c|} {Peak integral background rate [counts/day]} \\
\,[keV] &	&Gator (Soudan)	&  Gator (LNGS, 09-2007)	& Gator (LNGS, 10-2008)	&  Gator (LNGS, 04-2010)	& GeMPI  \cite{neder2000} \\
\hline
239	 &	$^{232}$Th/$^{212}$Pb				& 1.1   $\pm$ 0.7		& 0.7 $\pm$ 0.1 	&	0.13 $\pm$ 0.08  	&$<$0.5	 		& NA \\
911 	 &	$^{232}$Th/$^{228}$Ac				& 0.9   $\pm$ 0.3		& 0.4 $\pm$ 0.2 	&	0.4 $\pm$ 0.1 		&$<$0.5  			& $<$0.2 \\
352 	 & 	$^{238}$U/$^{214}$Pb				& 4.9	   $\pm$ 0.7		& 4.3 $\pm$ 0.7 	&	1.1 $\pm$ 0.2 		&0.7 $\pm$ 0.3	  	& $<$0.5 \\
609 	& 	$^{238}$U/$^{214}$Bi				& 4.5	   $\pm$ 0.5		& 4.0 $\pm$ 0.5  	&	1.1 $\pm$ 0.2 		&0.6 $\pm$ 0.2 		& 0.50$\pm$0.45 \\
1120	 & 	$^{238}$U/$^{214}$Bi				& 1.6   $\pm$ 0.3		& 2.7 $\pm$ 0.4  	&	1.3 $\pm$ 0.2 		&0.3	$\pm$ 0.1  	& NA \\
1765	 & 	$^{238}$U/$^{214}$Bi				& 1.3   $\pm$ 0.2		& 1.5 $\pm$ 0.3  	&	0.2 $\pm$ 0.1 		&0.08 $\pm$ 0.06  	& NA \\
662 	 &	$^{137}$Cs						& 2.9   $\pm$ 0.4		& 0.5 $\pm$ 0.3	&	$<$0.5 			&0.3 $\pm$ 0.1		& NA \\
1173	 & 	$^{60}$Co						& 0.5   $\pm$ 0.1		& 0.5 $\pm$ 0.2  	&	0.5 $\pm$ 0.2  		&0.5 $\pm$ 0.1 		& 0.6$\pm$0.4 \\
1332	 &	$^{60}$Co						& 0.6   $\pm$ 0.1		& 0.6 $\pm$ 0.2  	&	0.5 $\pm$ 0.1 		&0.5	$\pm$ 0.1  	& 0.4$\pm$0.3 \\
1461 & 	$^{40}$K							& 5.8   $\pm$ 0.4		& 0.5 $\pm$ 0.2  	&	0.4 $\pm$ 0.1  		&0.5	$\pm$ 0.1 		& 0.6$\pm$0.4 \\
2615 & 	$^{208}$Tl						& 0.7   $\pm$ 0.1		& 0.2 $\pm$ 0.1  	&	0.2 $\pm$ 0.1  		&0.2 $\pm$ 0.1 	& NA \\
\hline
\end{tabular}
\end{center}
\end{table}

\begin{figure}[!h]
\begin{center}
\includegraphics[scale=0.42]{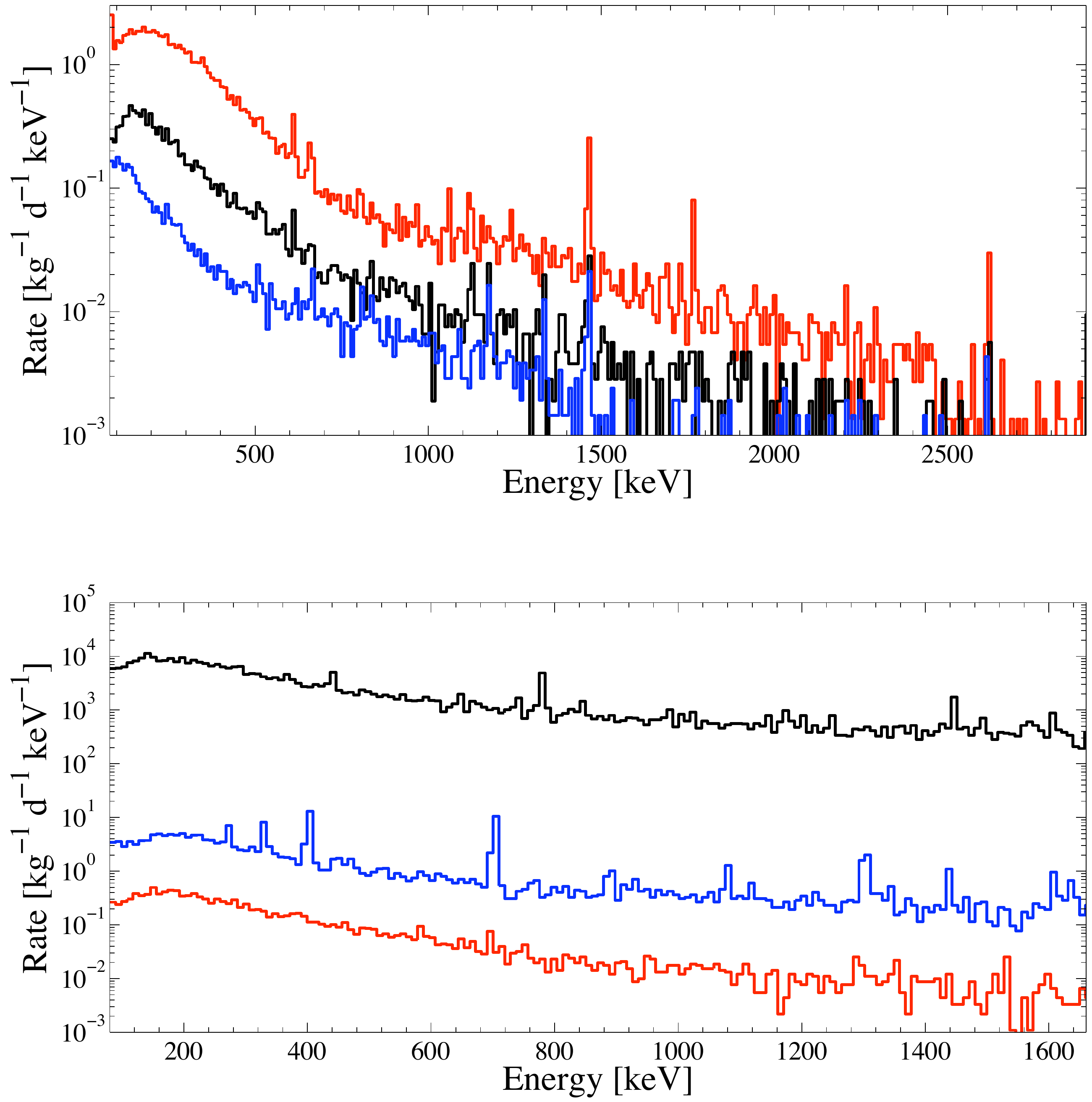}
\caption{\small{ (Top) Background spectra of Gator at Soudan (red), at
LNGS (black) and the spectrum of the GeMPI detector \,\cite{neder2000} (blue).  (Bottom) Gator background spectrum at LNGS: outside the shield (black), 
inside the shield (blue) and inside the shield with the radon protection system on (red), clearly showing the suppression of the main gamma lines associated with radon decays.}}
\end{center}
\label{fig:BG}
\end{figure}

Figure~\ref{fig:BG} (top) shows the comparison between the latest  background spectrum acquired at LNGS (2010), a spectrum taken in the SOLO facility at Soudan (2007) and 
the background of the GeMPI detector \,\cite{neder2000}.  It also shows (bottom) the Gator spectrum underground at LNGS prior to its installation in the shield, inside the shield, and inside the shield with the radon protection system on. 
The background decrease is more than four orders of magnitude, and shows that  a careful shielding is needed even when the detector is operated deep underground.

\section{Calibration measurements and efficiency determination}
\label{sec:efficiency}

\begin{figure}[!h]
\includegraphics[scale=0.33]{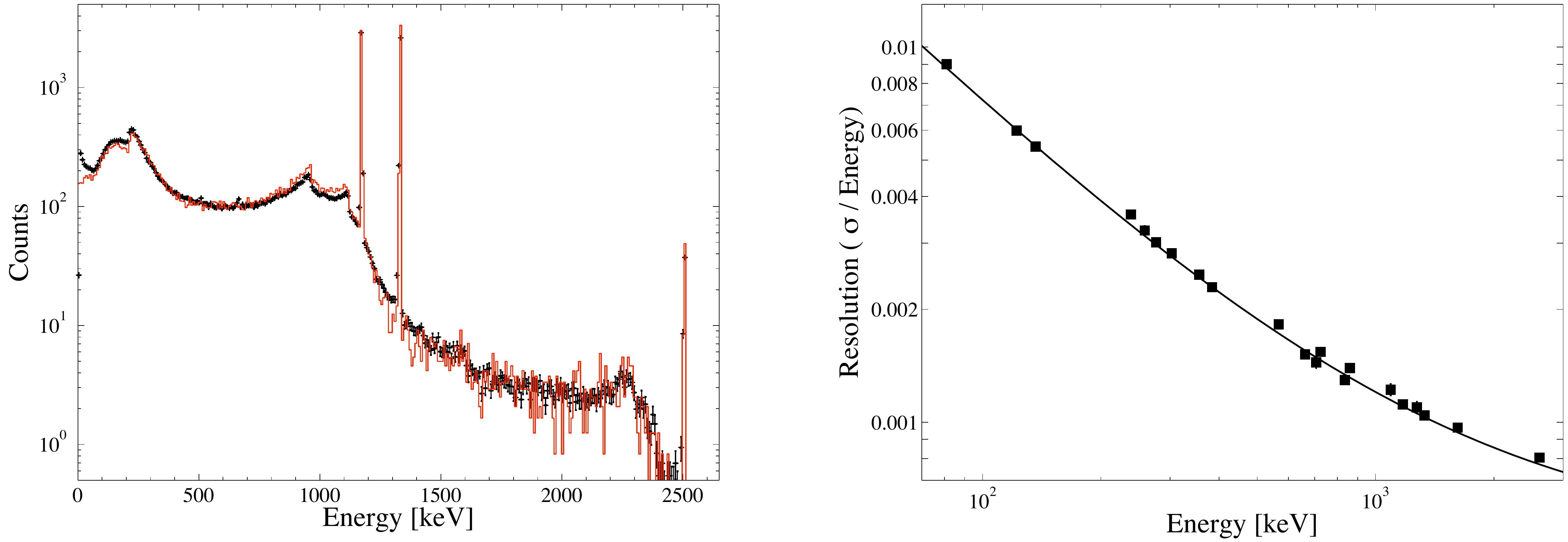}
\caption{\small{(Left) Comparison of the Gator spectrum from a $^{60}$Co calibration (black, data points) with the 
one from a Monte Carlo simulation of the source-detector geometry (red, solid). (Right) Energy resolution (defined here as the ratio of $\sigma$/energy) as a function of energy. 
The solid curves represent a fit using the function $\sigma^2(E) = E^2(2.35\times10^{-7}) + E(7.70\times10^{-4}) + (4.43\times10^{-1})$. }}
\label{fig:eresol}
\end{figure}

The HPGe detector is calibrated regularly with  radioactive sources such as  $^{109}$Cd, $^{133}$Ba, $^{137}$Cs, $^{60}$Co, $^{57}$Co, $^{22}$Na, $^{54}$Mn and  $^{228}$Th.
In Figure \ref{fig:eresol} (left) the comparison of the spectrum obtained from a $^{60}$Co calibration with the 
one from a Monte Carlo simulation of the source-detector geometry is shown. The FWHM of the two $^{60}$Co lines at 1173\,keV 
and 1332\,keV are 2.5\,keV and 3.0\,keV, respectively. The energy resolution of the detector, defined here as the ratio of the $\sigma$ to 
the mean energy of the gamma line, is shown in Figure \ref{fig:eresol} (right).

The calculation of the detection efficiencies of the various gamma lines used in the analysis of the experimental data (see Section \ref{sec:Analysis}) is based on Monte Carlo simulations using 
Geant4\,\cite{Agostinelli:2002hh}. 
For each measured sample, a detailed geometry is included into a Geant4 model of the facility. The efficiency $\epsilon$ of a specific $\gamma$-line  is
defined as the ratio between the number of events detected in the line to the number of gammas of that energy emitted by the source. In order to simulate
each decay chain, the G4RadioactiveDecay class, which takes into account the branching ratios for the different gamma lines in one decay, is used. 

\begin{figure}[!h]
\begin{center}
\includegraphics[scale=0.4]{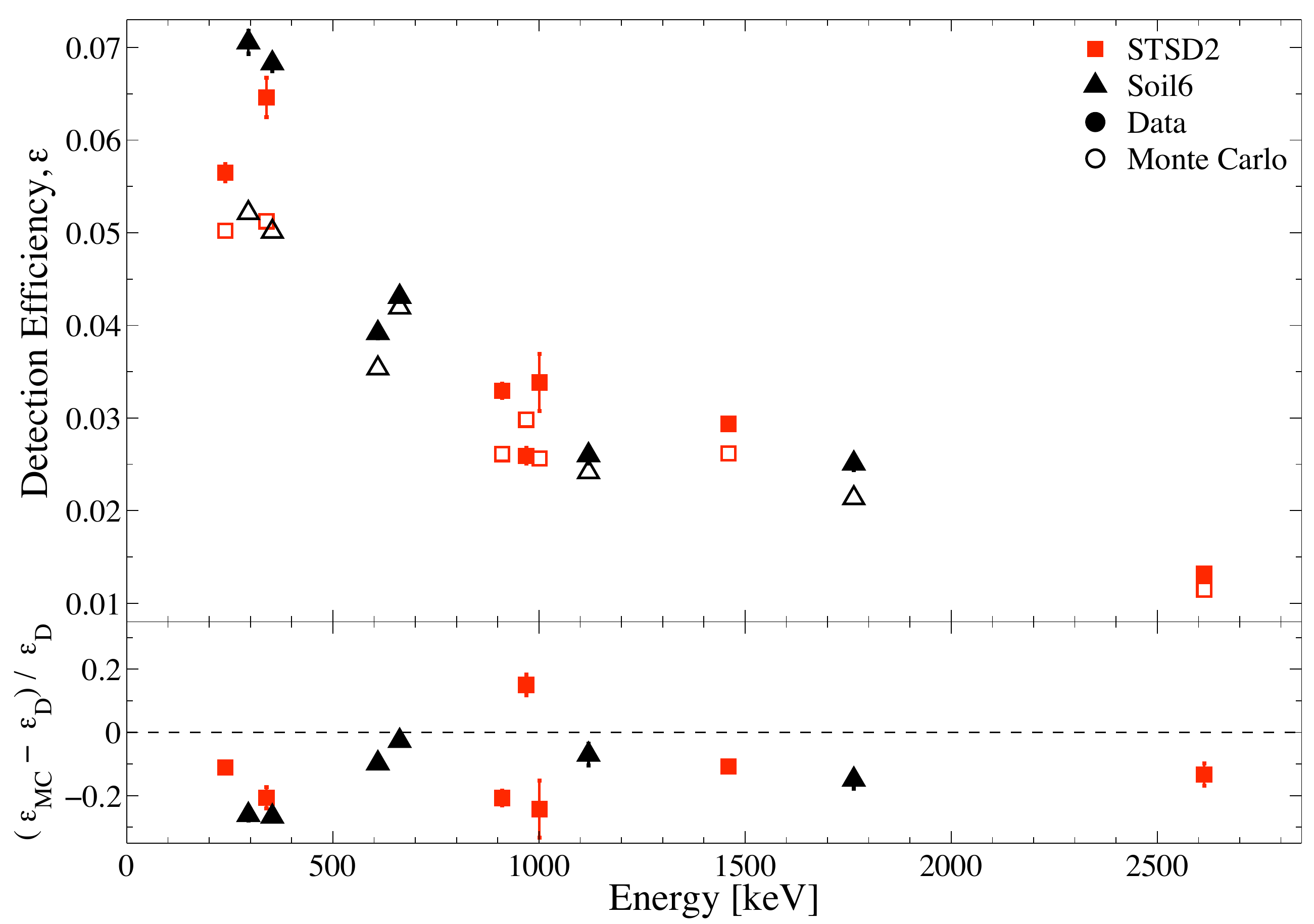}
\caption{\small{Upper panel: Efficiencies as a function of energy for the two certified sources, as determined from data (filled symbols) and Monte Carlo (open symbols). Lower panel: relative difference between simulated  (MC)  and measured (D) efficiencies.}}
\end{center}
\label{fig:certified}
\end{figure}

To cross-check the efficiency determination, a measurement of two extended sources 
and a comparison with the certified values for their activities has been performed.
The sources used for this measurements, which had similar dimensions and weights, are CANMET-STSD2 (from the Canada Centre for
Mineral and Energy Technology) and IAEA-Soil6 (from the International Atomic Energy
Agency). Both sources are soils from different places on Earth, which have been 
thermally treated and sieved several times in order to destroy any remaining organic
matter. The homogeneity of the material is  certified by the provider.  
Tables \ref{tab:stsd2} shows the results Gator's screening of these two sources  and a 
comparison with the certified values. Within the uncertainties of the measurements, we find a good agreement. We also show in Figure \ref{fig:certified} the 
comparison of the efficiencies as determined by the Monte Carlo method, with those from the data, as a function of energy, as well as the relative difference among these.

These results indicate that the measurements performed with our spectrometer provide a reliable value for the  activity of a given sample. Further cross-check were done by using 
well-calibrated, commercially available point sources.

\begin{table}[h]
\begin{center}
\caption{\label{tab:stsd2} Results from screening the CANMET-STSD2 (493\,g) and IAEA Soil6 (530\,g) sources and comparison with certified values provided by two agencies (for details, see text).}
\vspace{0.3cm}
\begin{tabular}{|l|rr|}
\hline
\multirow{2}{*} {Nuclide}	& \multicolumn{2}{c|} {STSD2 Activity [Bq/kg]} \\
	& Gator results	& Certified values	  \\
\hline
$^{228}$Th	& 75$\pm$4		& 70$\pm$5	  \\
$^{226}$Ra	& 230$\pm$30 &  230$\pm$10	 \\
$^{40}$K	        & 590$\pm$10 &  540$\pm$20	\\
\hline
\multirow{2}{*} {Nuclide}	& \multicolumn{2}{c|} {Soil6 Activity [Bq/kg]} \\
	& Gator results	& Certified values \\
\hline
$^{226}$Ra		& 88$\pm$5 	& 80$\pm$7	\\
$^{137}$Cs	& 57$\pm$2	& 54$\pm$2	\\
\hline
\end{tabular}
\end{center}
\end{table}

\section{Data analysis methods}
\label{sec:Analysis}

Two methods are used to determine the  concentrations of  radioactive nuclides in a given sample.  In the first method, the most prominent $\gamma$-lines  are analyzed, using  efficiencies as determined by a full Monte Carlo simulation. In the case of $^{238}$U, the $\gamma$-lines from the daughters of $^{226}$Ra ($^{214}$Pb and $^{214}$Bi) and  in  case of $^{232}$Th the $\gamma$-lines from $^{228}$Ac and from $^{212}$Pb, $^{212}$Bi and $^{208}$Tl are used.  
In the second method, the overall data spectrum is compared to the one obtained from a Monte Carlo simulation, after subtracting the measured background spectrum, and the activities are determined from the best fit.  The results from the two methods agree within the statistical errors, as shown in Section\,\ref{sec:Results}.  Both methods are explained in more detail below.

\subsection{Analysis of  gamma-lines}
\label{sec:gamma-lines}

The first method is based on counting events at the location of the most prominent lines,  after subtracting the background spectrum  closest in time. The Compton background, estimated from the regions left and right of  a peak, is subtracted as well. The decision on whether the signal exceeds the background is based on 
comparing the net signal number of counts 
\begin{equation}
S_{net} = S - B\cdot t_S/t_B - B_C 
\end{equation}
with the so-called detection limit $L_d$  (the level of a true net signal that can be detected with a given probability) as defined in \cite{Hurtgen200045} for a $\sim$95\% C.L.:
\begin{equation}
\label{ld}
L_d = 2.86 + 4.78 \sqrt{B_C + B\cdot \frac{t_S}{t_B} +1.36}.
\end{equation}
$S$ is the number of counts in the $\pm3\sigma$-region around a peak, $B$ and $B_C$ are  the number of background and Compton-background counts in the same region, and $t_S$, $t_B$ are 
the measuring times for signal and background, respectively. For each peak, three cases are considered\,\cite{Hurtgen200045}:
\begin{enumerate}
	\item $S_{net}<0$: the upper limit is set to $L_d$ (no net contribution from a signal)
	\item $0<S_{net}<L_d$: the upper limit is set to  $S_{net}+L_d$ (there is an indication of a signal, but it can not be confirmed for the existing background level and sample exposure)
	\item $S_{net}>L_d$: the detection limit is exceeded  (clear indication for a signal at 95\% C.L.)
\end{enumerate}

For the third case, the specific activity and its 1 $\sigma$ error is  calculated as
\begin{equation}
\label{eq:activity}
A {\rm {[Bq/kg]}} = \frac{S_{net}}{r \cdot \epsilon \cdot m \cdot t}   \quad \textnormal{with} \quad \frac{\Delta A}{A} = \frac{\Delta S_{net}}{S_{net}}.
\end{equation}
with the peak  efficiencies $\epsilon$ as determined by Monte Carlo simulations, the branching ratio $r$ for the specific line, the mass $m$ of the sample (in kg), and the measuring time $t$ (in seconds).
For the  case in which an upper limit is reported, $S_{net}$ is replaced by $L_d$ or by ($S_{net} + L_d$) in  equation (\ref{eq:activity}).
 As concrete examples, we show the above quantities in Table \ref{tab:limits}, together with the determined specific activities or upper limits for a copper and a stainless steel sample,
using different gamma lines from their measured spectrum.

\begin{table}[h]
\begin{center}
\caption{\label{tab:limits}{Examples of upper limit or specific activities calculation. Details are given in the text.}}
\vspace{0.3cm}
\begin{tabular}{|c|ccccccc|}
\hline
 Sample	& Used line [keV] &	$S_{net}$	&	$B\cdot t_S/t_B$	&	$B_c$	&	$L_d$	&	Condition & Activity  [mBq/kg] \\
 \hline
Copper 		& 239		 & 0			&	0	&	93	&	49		&	$S_{net}<L_d$  & $<$0.33 \\
			& 352		 & -5			&	19	&	71	&	48		&	$S_{net}<L_d$  & $<0.36$\\
			& 1173		 & 42			&	6	&	19	&	27		&	$S_{net}>L_d$ & 0.24$\pm$0.06\\
	\hline
Stainless steel	& 352		 & 66			&	7	&	58	&	42		&	$S_{net}>L_d$ & 4.3$\pm$0.9 \\
			& 1173		 & 236		&	2	&	19	&	25		&	$S_{net}>L_d$  & 7.2$\pm$0.9\\
			& 1461		 & -3			&	3	&	10	&	21		&	$S_{net}<L_d$  & $<$5.7\\
\hline
\end{tabular}
\end{center}
\end{table}

\subsection{Fit of the data to a simulated spectrum}
\label{sec:chi-sq}

The individual activities are also determined by a global fit based on a $\chi^2$-minimization of the simulated spectrum to the experimental data.
The aim is to model $\vec{y}$, the measured number of counts in each energy bin, as a function of $\vec{x}$, the
energy bin value from the Monte Carlo simulation, with a functional
dependence of the form
\begin{equation}
\vec{y} = \sum_{k=0}^{M} a_k \cdot f_k(\vec{x}),
\label{eq:chi2}
\end{equation}
where $M$ is the number of simulated radioactive isotopes. $a_{k}\geq0$ are the scaling factors for each isotope, which are kept as free
parameters,   and $f_{k}(\vec{x})$ are $M$ Monte Carlo spectra which are already smeared with the energy resolution of the detector.
For uncorrelated statistical errors in each energy bin $i$, the $\chi^2$
value is defined as
\begin{equation}
\chi^2 = \sum_{i=0}^{N} \frac{\left[y_i - \sum_{k=0}^{M}a_k\cdot f_k(x_i)\right]^{2}}{\sigma_i^{2}},
\label{eq:chi2}
\end{equation}
where $\sigma_i = \sqrt{y_i}$ is the variance in the observed number of
counts in each bin and $N$ is the number of energy bins over which the
fit is performed. The statistical uncertainty in the Monte Carlo
component is negligible. Before the analysis, the measured background spectrum is subtracted from
the sample spectrum. The outcome of this procedure are the scaling factors $a_k \pm \Delta
a_k$ for every decay chain/isotope $k$.  In the subsequent analysis, these factors are assumed to describe a normalized Gaussian probability
distribution function (PDF) 
with mean $a_0=a_k$ and $\sigma = \Delta a_k$.

The  method applied to decide between a real detection and an upper limit is based on \cite{edoc:451164}. If the lower of the two symmetric
limits providing 95\% statistical coverage is positive, a peak detection is claimed and the activity is calculated from $a_k$, taking into
account the measuring time $t$ in seconds, the sample mass $m$ in kg, and the number of simulated events $n_{sim}$.
The relative $\pm 1 \sigma$ error is given by $\Delta a_k/a_k$:
\begin{equation}
\label{eq:limits_chi2}
A_k {\rm {[Bq/kg]}}= \frac{n_{sim} \ a_k}{m \ t} \quad \textnormal{with} \quad \frac{\Delta A_k}{A_k} = \frac{\Delta a_k}{a_k}.
\end{equation}
If the lower limit is equal or less than zero, no detection can be
claimed at 95\% C.L. and  an upper limit is given. Its value $a_{up}$ is
determined by the 95\% quantile of the positive part of the Gaussian PDF
defined by $a_0=a_k$  and $\sigma = \Delta a_k$. The limit on the activity
(95\% C.L.) is calculated using $a_{up}$ in equation (\ref{eq:limits_chi2}).

\begin{figure}[!h]
\begin{center}
\includegraphics[scale=0.45]{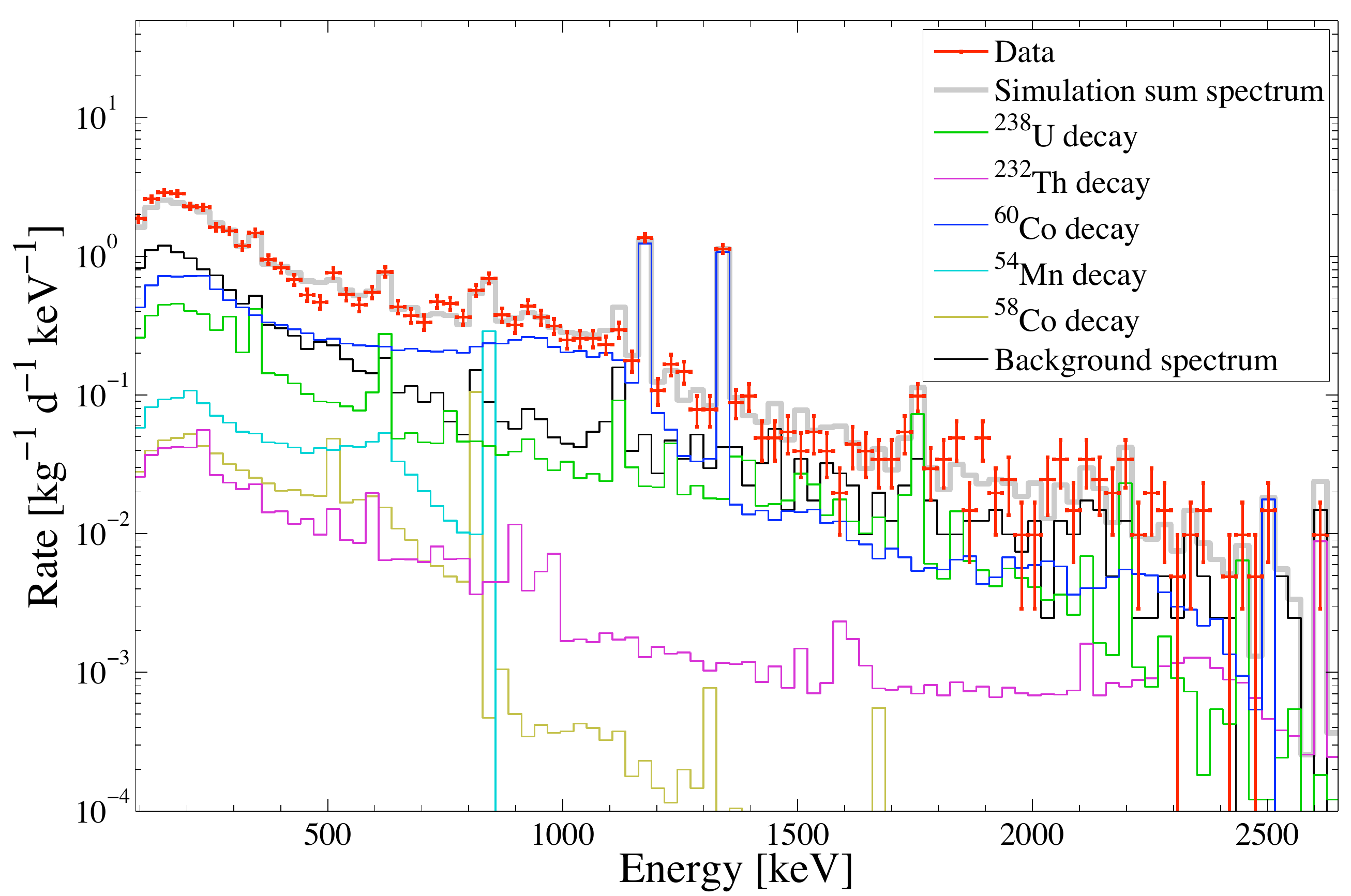}
\caption{\small{Best-fit of the Monte Carlo simulations to the measured
data for a  stainless steel sample. Data is shown with error bars (red), the full spectrum as determined by the simulation is shown as the grey 
solid curve. The individual contributions from the decays of  $^{60}$Co (blue),
$^{238}$U (green), $^{232}$Th (magenta), $^{54}$Mn (cyan) and $^{58}$Co (brown) are also shown, along with the measured background spectrum (black).}}
\end{center}
\label{fig:3mmSS_fit}
\end{figure}

As an example,  Figure \ref{fig:3mmSS_fit} shows a measurement of a stainless steel sample: the data spectrum is compared  with the Monte Carlo sum spectrum, and the individual contributions from $^{238}$U, $^{232}$Th, $^{40}$K,
and $^{60}$Co, as given by the best-fit, are shown. The derived activities and upper limits for this sample are given in Table  \ref{tab:comparison}.

\section{Results}
\label{sec:Results}

In this section, we first outline results obtained from a detailed study of the background of the facility, using the analysis method described in Section \ref{sec:chi-sq}. We then present screening results for a few selected samples, the goal being to compare the outcome of the two analysis methods introduced in the previous section.  Results from a much larger selection of screened samples are given and discussed in detail in \cite{Screening}.

\subsection{Background Analysis}
\label{sec:Backgrounds}

To model the residual background of the Gator facility, the detailed geometry of the crystal, cryostat system and shields  has been simulated with Geant4.
The following potential contributions to the background have been simulated:  the natural decay chains of $^{238}$U, $^{232}$Th and $^{40}$K decays in the copper 
of the shield and of the cryostat,  the decays of the cosmogenic radio-nuclides $^{54}$Mn, $^{65}$Zn,$^{58}$Co, $^{57}$Co, $^{60}$Co in the Ge crystal, 
the Cu of the cryostat and of the shield, $^{210}$Pb decays in the innermost Pb layer, and $^{222}$Rn decays inside the shield.\footnote{A background run up to an energy of $\sim$8\,MeV was 
acquired for a few weeks and revealed no U/Th contaminations close to the crystal, which would be visible as alpha peaks above a few MeV.} Other materials in the cryostat (PTFE, mylar and kapton) were neglected in the simulation due to their much lower mass compared to the copper (a few tens of grams and milligrams, respectively).

\begin{figure}[!h]
\begin{center}
\includegraphics[scale=0.40]{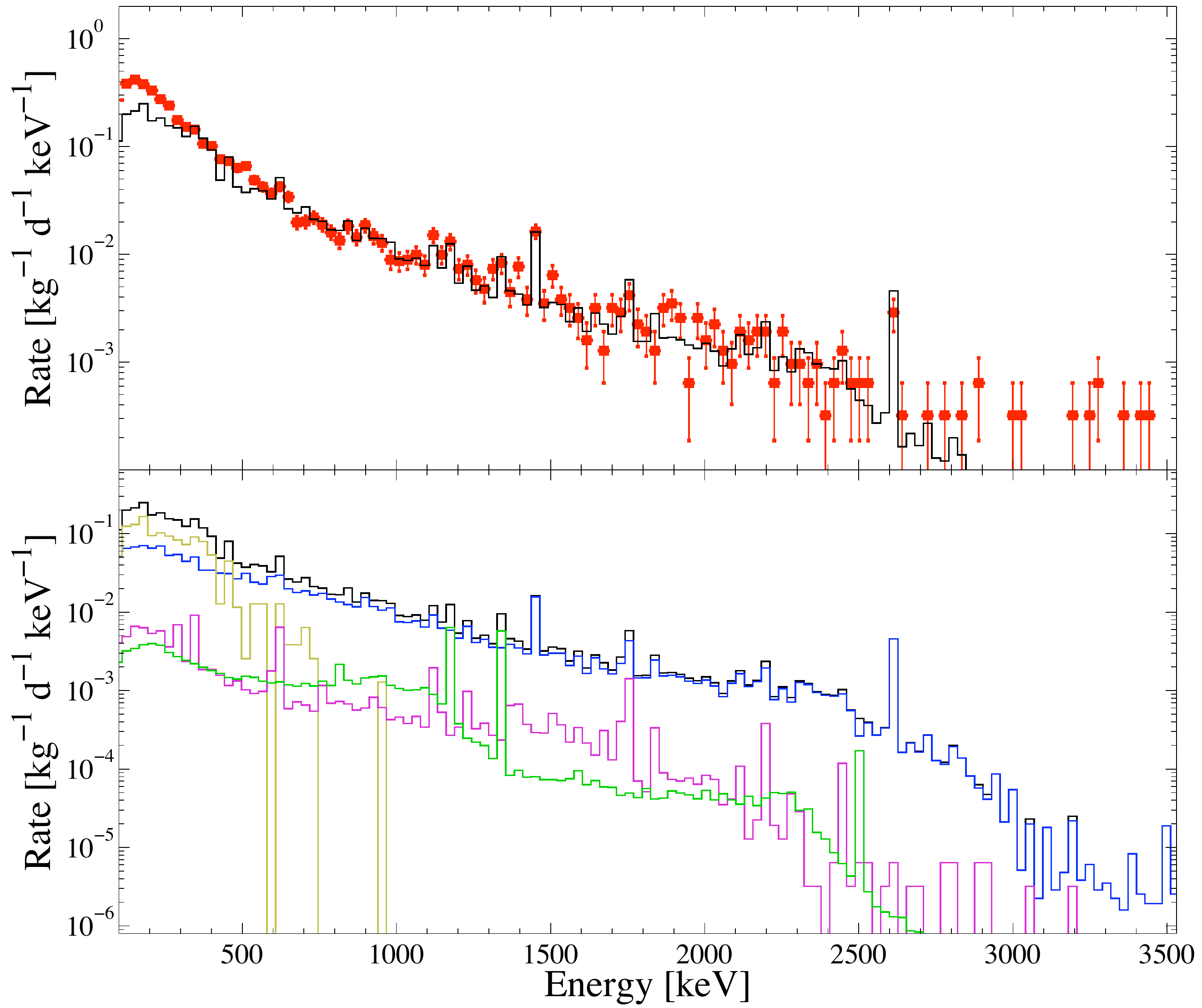}
\end{center}
\caption{\small{(Top) Comparison of the sum of the simulated spectra from natural
and cosmogenic radionuclides in the detector and shield materials (black) with the observed background spectrum (red data points). 
(Bottom) The individual, best-fit contributions to the observed spectrum are shown: natural radioactivity in Cu (blue),  cosmogenic radio-nuclides in  
Ge and Cu (green), $^{222}$Rn decays inside the shield (magenta) and $^{210}$Pb decays in the Pb shield (yellow).}}
\label{fig:BG3}
\end{figure}

Figure \ref{fig:BG3} (top) shows the comparison of the latest measured Gator background to the sum spectrum obtained from the Monte Carlo simulations. 
On the bottom, the individual contributions of the main simulated background components are  shown. The resulting activities for the individual chains and 
radio-nuclides,  using the best fit of the data to the Monte Carlo simulations, are presented in Table \ref{tab:background}. The background of the detector above $\sim$500\,keV is thus currently dominated by the residual $^{238}$U, $^{232}$Th, $^{40}$K and $^{60}$Co activities in the Cu of the shield, and, to a lower extent, in the copper of the cryostat. Below $\sim$500\,keV, the background is dominated by $^{210}$Bi bremsstrahlung from $^{210}$Pb decays in the innermost lead shield. The obtained $^{210}$Pb activity is with 5.7$\pm$0.5 Bq/kg higher than 3 Bq/kg, which is  the value specified by the provider.

\begin{table}[h]
\begin{center}
\caption{\label{tab:background}\small{Activities of residual and cosmogenic
radionuclides inside the copper shield, the Ge crystal and the cryostat and of $^{222}$Rn inside the shield.}}
\vspace{0.3cm}
\begin{tabular}{|l|ccccc|}
\hline
Isotope &                \multicolumn{3}{c}{Specific activity [$\mu$Bq/kg]}   & Activity  [$\mu$Bq/m$^3$] & Activity [Bq/kg]\\ 
        (chain)                           &            Cu (shield)           & Ge crystal & Cu (cryostat)  & Sample chamber & Inner Pb shield\\
        \hline
{$^{226}$Ra ($^{238}$U) }                 &                                56 $\pm$ 11                      	     &                                                           &                           8 $\pm$ 5         &           &            \\  
{$^{228}$Th ($^{232}$Th)}              &        27 $\pm$ 7                               &                                                         &                           4 $\pm$ 2                 &               & \\    
{$^{40}$K }                   &                  		32 $\pm$ 13                             &                  $<$1.30                                 &                            11 $\pm$ 6             &                &   \\  
{$^{60}$Co }                 &                 		8 $\pm$ 4                                 &                 $<$0.80                                 &                          1.3 $\pm$ 0.4                   &             &   \\   
{$^{58}$Co }                 &               		 $<$0.27                                      &                 $<$0.11                                 &                           $<$0.22                             &        & \\   
{$^{54}$Mn }                 &                		$<$1.30                                      &                 $<$1.60                                  &                           $<$2.15                               &     &   \\  
{$^{65}$Zn }                   &               		 $<$0.16                                      &                  $<$0.15                                 &                           $<$0.50                                 &    &   \\  
{$^{222}$Rn ($^{238}$U)}                   &               				                                  &             	                           &                          	                        &    $<$55   &  \\  
{$^{210}$Pb}                   &               				                                  &             	                           &                          	                        &      & 5.7 $\pm$ 0.5  \\  
\hline
\end{tabular}
\end{center}
\end{table} 
 
\subsection{Sample Analysis}
\label{sec:Samples}

\begin{figure}[!h]
\includegraphics[scale=0.3]{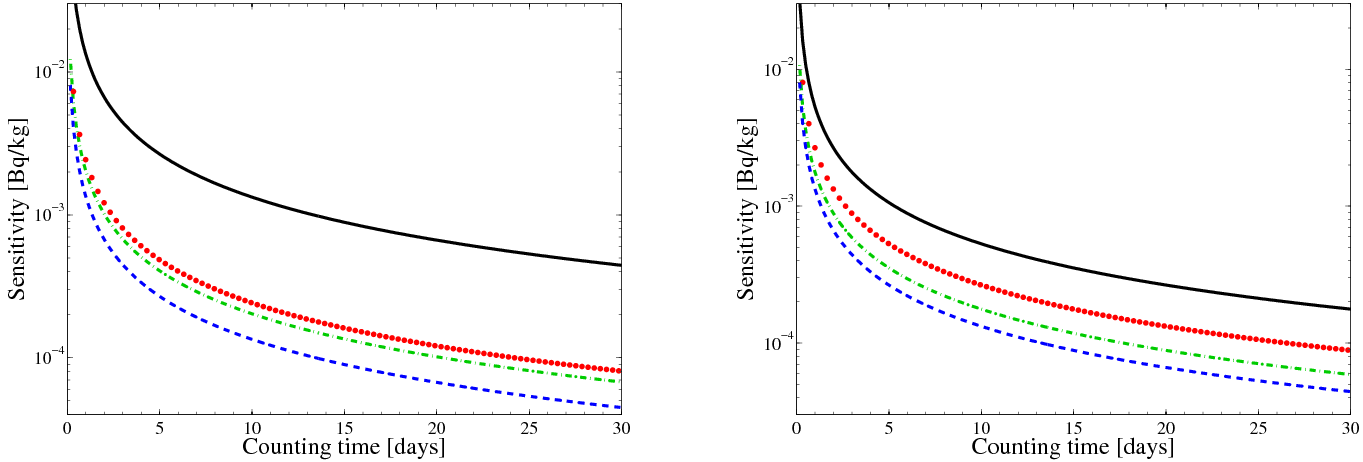}
\caption{\small{(Left) Sensitivity versus time for $^{40}$K (black),  the $^{238}$U chain (using the $^{214}$Bi 609\,keV line, red),  the  $^{232}$Th chain (using the $^{212}$Pb 239\,keV line, green) and $^{60}$Co  (blue) for a 13.5\,kg sample of PTFE.
(Right)  Sensitivity versus counting time for different PTFE sample masses: 4\,kg (black), 8\,kg (red), 12\,kg (green) and 16\,kg (blue) using the 609\,keV gamma line from $^{214}$Bi. }}
\label{fig:sens_time}
\end{figure}

Prior to the screening of a specific sample, an estimate of the minimal measuring time $t_{min}$, based on the mass, shape and the targeted activity $A_{target}$ is performed:
\begin{equation}
t_{min} = \frac{L_d}{A_{target} \cdot r \cdot m \cdot \epsilon}, 
\end{equation}
where $m$ is the mass of the sample, $r$ is the branching ratio, $\epsilon$ is the detection efficiency and the detection limit $L_d$ is defined in equation (\ref{ld}).
Typical numbers for $t_{min}$ are 1-3 weeks in order to achieve a sensitivity below a few mBq/kg.  Figure \ref{fig:sens_time} shows the sensitivity versus counting time for different sample masses using a specific gamma line (right) and for various radio-isotopes for a specific sample  (left).

The activities of the screened materials  are calculated using the analysis methods  described in Section \ref{sec:gamma-lines} and  \ref{sec:chi-sq}.  As evident from Table \ref{tab:comparison}, where a few results from  screened samples are provided, an excellent agreement between the results of the two methods was achieved.  It provides a cross-check on the reliability of the used methods and on the obtained activities (or upper limits) for  the samples that were screened. A large set of measurements is described and discussed in detail in\,\cite{Screening}. 

\begin{table}[h!]
\begin{center}
\caption{\label{tab:comparison}{Comparison of results obtained by the two different analysis methods using four different screening samples.}}
\vspace{0.3cm}
\begin{tabular}{|l|cccccc|}
\hline
{Material}                      	&              {Method}   	  		 & Activity        &                {$^{226}$Ra }      	&                {$^{228}$Th }        	 &      {$^{40}$K }         &          {$^{60}$Co }                    \\  
(amount, time)	& & & & & & \\
\hline 
{Copper}               		&             $\chi ^{2}$  			&   mBq/kg       &                 $<$0.5                  	 &                 	$<$0.5           	&       $<$0.9      		&        0.24  $\pm$0.08            \\  
(18\,kg, 20\,d)					&	  $\gamma$-lines	 	&    mBq/kg      &		$<$0.3 	       		 &	        		$<$0.3           	 &      $<$1.3       		 &	 0.24  $\pm$0.06	  \\ 
					\hline 

{Stainless steel }              	&              $\chi ^{2}$     		&  mBq/kg     	&             	4.1$\pm$0.6        	 &                   	 1.4$\pm$0.4         	 &  	$<$4.6                      &          7.4$\pm$1.1                  \\   
(6.6\,kg, 6.8\,d)					&	      $\gamma$-lines		&  mBq/kg      	&		4.3$\pm$0.9	       	&		      	$<$1.8	 	&	 $<$5.7                     &		7.2$\pm$0.9		 \\ \hline 

{Concrete }     	          	&              $\chi ^{2}$         		& Bq/kg       	&    		17$\pm$2     	    & 	       		3.1$\pm$0.7         	 &      53$\pm$4 	         &        $<$0.6                  	\\   
(35\,g, 0.7\,d)					&	 $\gamma$-lines		& Bq/kg 	 	&		15$\pm$2	    	   &	      		 3.8$\pm$0.8		&	42$\pm$6     		&	$<$0.7   		\\ 
\hline 

{Photomultipliers }              &               $\chi ^{2}$        		&  mBq/PMT  	&                  $<$0.2                  	&           	0.20$\pm$0.09                      &  	8$\pm$1        &		0.6$\pm$0.1		\\  
(22 pieces, 5.5\,d)					&	   $\gamma$-lines		& mBq/PMT  	&  		  $<$0.2 			&		0.18$\pm$0.06			&     11$\pm$2          &		0.6$\pm$0.1			    \\ \hline

\end{tabular}
\end{center}
 \end{table}

\section{Summary}

The Gator screening facility at LNGS, described in detail in this paper, includes one of the world's highest sensitivity HPGe spectrometers. It allows to measure large samples and to reach a sensitivity on  specific activities below $\sim$mBq/kg when using typical sample masses of a few kg and measuring times extending from one to several weeks. 
The integral counting rate of the detector in the 100-2700\,keV energy region has decreased by more than a factor of 5 from (0.842$\pm$0.005) counts/min at Soudan to (0.157$\pm$0.001)  counts/min 
at LNGS. This  decrease is mainly due to the improved shield and radon protection system, but also because of the decay of cosmogenic isotopes in the crystal and surrounding copper. 
The background of the facility, which was modeled using a detailed geometry of the spectrometer and its shield, and Monte Carlo simulations with  Geant4,  is dominated by U/Th/K decays in the copper of the cryostat and the shield, and from residual $^{210}$Pb in the innermost lead shield.  This results  from a comparison of background measurements with the simulations, by choosing the best fit of the MC spectra to the measured data.
Two different data analysis methods show an excellent agreement when applied on a variety of samples.  Detailed screening results for the XENON100 \cite{Aprile:2010um} experiment are presented and discussed in\,\cite{Screening}. 
At present, the Gator facility is operated under stable conditions at LNGS. It is used to screen  materials for the construction of the XENON1T experiment,  for the next phase of the GERDA project, 
as well as for a future noble liquid dark matter search facility, DARWIN\,\cite{Baudis:2010ch}.

\newpage

\section{Acknowledgements}

We thank James Beaty from the University of Minnesota for help with the detector operation at the Soudan laboratory and the machine shop crew at the RWTH Aachen, in particular 
Dipl. Ing. Michael Wlochal for the collaboration in the design and construction of the shield. 
We thank Giuseppina Mosca and Stefano Nisi  from the chemistry lab at LNGS for the assistance in cleaning the samples, the LNGS staff,  in particular Ing. Piergiorgio Aprili,
for their continuous support and the XENON collaboration for the help in maintaining the facility. This work is supported by the Volkswagen Foundation, by the University of Zurich, and 
by the Swiss National Foundation Grant No. 20-118119 and No. 20-126993. 

\bibliographystyle{plain}

\end{document}